\documentclass[twocolumn]{aastex631}

\usepackage[caption=false]{subfig}
\usepackage{float}
\usepackage{graphicx}	
\usepackage{amsmath}	
\usepackage{amssymb}
\usepackage{xcolor}
\usepackage{booktabs}
\usepackage{enumerate}
\usepackage{dirtytalk}
\usepackage{float}
\usepackage{enumitem}
\usepackage{booktabs}
\usepackage{lipsum} 
\usepackage{multirow}
\usepackage{makecell}
\usepackage[english]{babel}
\usepackage{comment}
\usepackage{multirow}
\usepackage[para]{footmisc}
\usepackage{footnote}
\usepackage[normal]{threeparttable}

\accepted{24 December 2022, The Astrophysical Journal Letters}

\begin{document}

\title{Early Insights for Atmospheric Retrievals of Exoplanets using JWST Transit Spectroscopy}

\correspondingauthor{S. Constantinou, N. Madhusudhan}
\email{\\ sc938@cam.ac.uk, nmadhu@ast.cam.ac.uk}

\author{Savvas Constantinou}
\affiliation{Institute of Astronomy, University of Cambridge, Madingley Road, Cambridge CB3 0HA, UK}

\author{Nikku Madhusudhan}
\affiliation{Institute of Astronomy, University of Cambridge, Madingley Road, Cambridge CB3 0HA, UK}

\author{Siddharth Gandhi}
\affiliation{Leiden Observatory, Leiden University, Postbus 9513 2300 RA, Leiden, The Netherlands}
\affiliation{Department of Physics, University of Warwick, Coventry CV4 7AL, UK}
\affiliation{Centre for Exoplanets and Habitability, University of Warwick, Gibbet Hill Road, Coventry CV4 7AL, UK}

\begin{abstract}

We have entered the era of the James Webb Space Telescope (JWST). We use the first JWST transmission spectrum of the hot Saturn-mass exoplanet, WASP-39~b, obtained with the NIRSpec instrument in the 3-5~$\mu$m range to investigate (a) what atmospheric constraints are possible with JWST-quality data in this spectral range, (b) requirements for atmospheric models used in retrievals, (c) effect of differences between data reduction pipelines on retrieved atmospheric properties, and (d) complementarity between JWST data in the 3-5~$\mu$m range and HST observations at shorter wavelengths. JWST spectra in the 3-5~$\mu$m range provide a promising avenue for chemical detections while encompassing a window in cloud opacity for several prominent aerosols. We confirm recent inferences of CO$_2$, SO$_2$, H$_2$O, and CO in WASP-39~b, report tentative evidence for H$_2$S, and retrieve elemental abundances consistent with Saturn's metallicity. We retrieve molecular abundances with $\sim$0.3-0.6 dex precision with this relatively limited spectral range. When considering the 3-5~$\mu$m data alone, reported differences in spectra with different reduction pipelines can affect abundance estimates by up to $\sim$1 dex and the detectability of less prominent species. Complementing with data at shorter wavelengths, e.g. with other JWST instruments or HST WFC3 ($\sim$0.8-1.7~$\mu$m), can significantly improve the accuracy and precision of the abundance estimates. The high data quality enables constraints on aerosol properties, including their composition, modal size and extent, motivating their consideration in retrievals. Our results highlight the promise of JWST exoplanet spectroscopy, while underscoring the importance of robust data reduction and atmospheric retrieval approaches in the JWST era.
\end{abstract}


\section{Introduction} \label{sec:introduction}

The first observations with the James Webb Space Telescope (JWST) are now available, heralding the dawn of a new era in our understanding of exoplanetary atmospheres. With a virtually complete coverage of the near-mid infrared, transmission spectroscopy with JWST enables simultaneous constraints on multiple chemical species and other physical properties in exoplanetary atmospheres \citep{Beichman2014, Stevenson2016b, Batalha2017a, Kalirai2018, Bean2018, Sarkar2020}. The generational leap in our understanding of chemical and physical processes in exoplanets is already underway.

Exoplanet transmission spectroscopy with the Hubble Space Telescope (HST), along with ground-based observations with facilities like the Very Large Telescope (VLT), have over the last 20 years been a key driver of the field's remarkable growth \citep{Seager&Sasselov2000, Charbonneau2002, VidalMadjar2003, Deming2013, Ehrenreich2015, Sing2016, Nikolov2016, Kreidberg2018}. Paired with theoretical developments in atmospheric modelling and retrievals, HST transmission spectra in the optical and near-infrared (NIR) have led to important constraints on the abundances of chemical species including H$_2$O, Na and K, as well as the properties of clouds and hazes in several exoplanetary atmospheres \citep{Madhusudhan2009, Madhusudhan2014b, Kreidberg2014, Wakeford2018, Barstow2017, Pinhas2019, Welbanks2019b, Madhusudhan2018}. Most of these atmospheric detections were made for irradiated gas giants, whose relative rarity \citep{Howard2010, Mayor2011, Wang2015, Fulton2021} is offset by their comparative observability. This is largely due to their extended, hydrogen-dominated atmospheres giving rise to large spectral signatures and hence high signal-to-noise observations.

Our goal in this work is to obtain a first glimpse into atmospheric properties of exoplanets that can be retrieved with JWST-quality transmission spectra. While observations with HST have been limited to wavelengths below 1.7 $\mu$m, JWST promises a substantial increase in both sensitivity and spectral range. In particular, the $\sim$3-5 $\mu$m range accessible with the NIRSpec instrument \citep{Ferruit2012} opens uncharted territory in chemical discovery space, as evidenced by recent inferences of CO$_2$ and SO$_2$ in the atmosphere of an exoplanet \citep{ERS2022,Rustamkulov2022,Alderson2022}.
Here we assess atmospheric constraints that are possible with JWST transmission spectra in the $\sim$3-5$\mu$m range and modeling requirements for retrieval frameworks in the JWST-era. We further investigate the sensitivity of retrievals to differences in spectra obtained using different data reduction pipelines as well as complementarity with NIR spectra with the HST/WFC3 instrument (0.8-1.7~$\mu$m). 

We focus on WASP-39~b, which is one of the first exoplanets whose transmission spectrum has been observed with JWST. The planet has a mass of $0.28 \pm 0.03$~M$_\mathrm{J}$, a radius of $1.28 \pm 0.04$~R$_\mathrm{J}$ and a zero-albedo equilibrium temperature of 1170~K \citep{Faedi2011, Mancini2018}. It orbits a G8-type host star with an intermediate brightness of J = 10.7 and V = 12.1 \citep{Faedi2011}. WASP-39~b is therefore an example of the immense diversity in the known exoplanet population. While its closest solar system analogue by mass is Saturn (M = 0.3~M$_\mathrm{J}$), it is significantly larger, with a radius greater than Jupiter's and significantly more strongly irradiated than any solar system gas giant. The notably low gravity of WASP-39~b makes its atmosphere highly conducive to transmission spectroscopy observations and has already led to detections of H$_2$O, Na and K with prior HST and ground-based facilities \citep{Fischer2016, Nikolov2016, Sing2016, Tsiaras2018, Wakeford2018, Pinhas2019, Kirk2019, Welbanks2019b, Kawashima2021}. As a result, WASP-39~b is the target of choice for the JWST Early Release Science (ERS) program, which has already led to novel inferences of CO$_2$ and SO$_2$ in its atmosphere \citep{ERS2022, Alderson2022, Rustamkulov2022}.

We consider the JWST observations of WASP-39~b over the 3-5~$\mu$m range, obtained with the NIRSpec PRISM spectrograph\citep{ERS2022}. Beyond its high data quality, this spectral range is representative of the majority of observations JWST is set to make over Cycle 1. Specifically, most Cycle 1 observations are set to be made with the NIRSpec spectrograph using the G395 grating over a similar $\sim$3-5~$\mu$m range. As such, our work is also a feasibility study, with the present observations of WASP-39~b constituting a near-best case scenario.

We also consider how minor variations in the JWST observations, particularly those arising from differences between reduction pipelines, can affect the retrieved atmospheric constraints. We do this by retrieving on two different reduction of the same observations, obtained with the Tiberius and Eureka pipelines, which have been reported to give slightly different results, particularly over the 3.6~$\mu$m Spitzer band \citep{ERS2022}.

Besides analysing the JWST data alone, we also consider their complementarity with prior observations. For this work, we consider pairing the 3-5~$\mu$m JWST observations with those obtained previously with HST/WFC3  (0.8-1.7~$\mu$m), examining how this affects the precision and accuracy of atmospheric constraints. This pairing is particularly important, as several of the Cycle 1 targets have or are set to be also observed with HST/WFC3. In doing so, we seek to assess the complementarity between JWST 3-5~$\mu$m observations and HST. 

In what follows, we discuss the observations in section \ref{sec:observations} and our retrieval methodology in section \ref{sec:methods}. The results of our investigation are presented in section \ref{sec:wasp-39b}, followed by our summary and discussion in section \ref{sec:discussion}.

\section{Prior and Current Observations}\label{sec:observations}

 \begin{figure*}
    \centering

    \includegraphics[width=\textwidth]{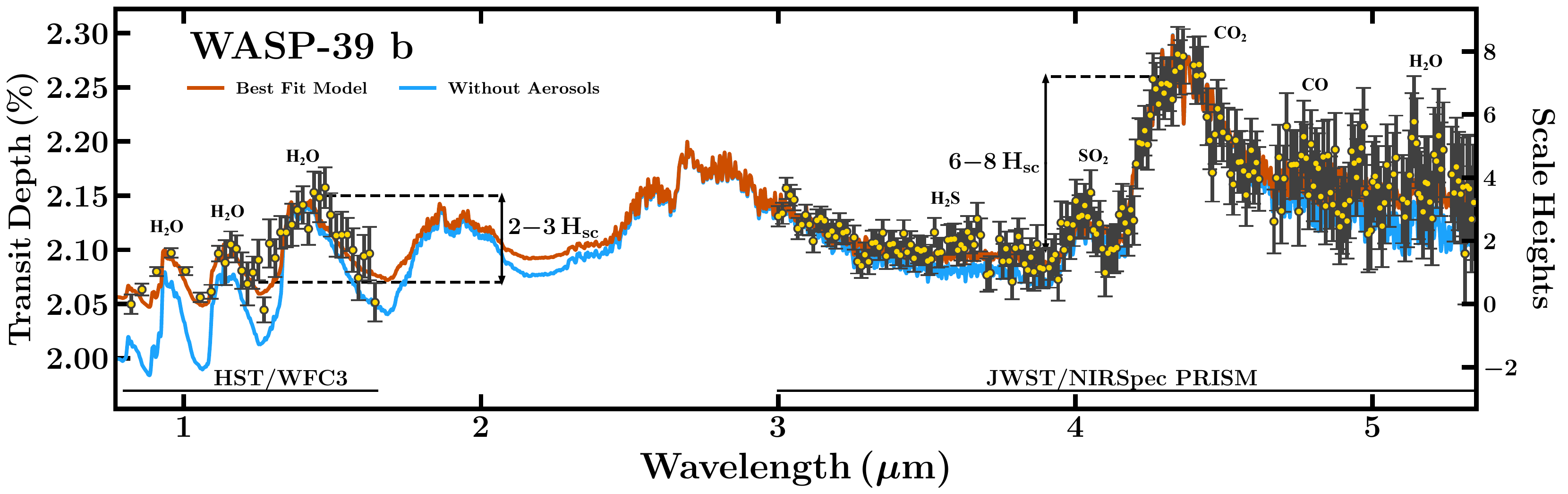}
    \caption{ A transmission spectrum of WASP-39~b. The circles with error bars show the JWST NIRSpec PRISM spectrum in the 3-5~$\mu$m range reduced with the Tiberius pipeline \citep{ERS2022} along with prior HST/WFC3 observations in the 0.8-1.7 $\mu$m range \citep{Wakeford2018}. The solid curve in brown shows our retrieved best fit spectrum, and the same spectrum but without opacity contributions from aerosols is shown in blue for reference (see section~\ref{subsubsubsec:tiberius+wfc3}). The heights of the prominent spectral features in the JWST and HST bands in terms of a characteristic atmospheric scale height are denoted by arrows; a nominal slant photospheric temperature of 800 K is assumed motivated by the retrieved constraints. The contributions of individual molecules are shown in figure~\ref{fig:contributions} in the Appendix.}   
    \label{fig:bestfit}
\end{figure*}

The atmosphere of WASP-39~b has been extensively probed in transmission spectroscopy from both ground and space. The first space-based observations were carried out in 2013, using the HST/STIS G430L and G750L gratings (GO 12473, PI: D Sing) \citep{Fischer2016, Sing2016}, covering the 0.29-1.0~$\mu$m wavelength range. Paired with photometric observations with Spitzer (90092, PI: J.-M. D{\'e}sert), these observations were used to infer a cloud-free atmosphere, with prominent spectral features arising from Na and K. This is in agreement with conclusions drawn from ground-based VLT observations by \citet{Nikolov2016}, spanning the 0.4-0.8~$\mu$m range using the Focal Reducer/Low Dispersion Spectrograph 2 (FORS2) (096.C-0765, PI: N. Nikolov).

Further spectroscopic observations were carried out in the NIR using the the HST Wide Field Camera 3 (WFC3) G102 and G141 grisms (GO 14169, PI: H. Wakeford; GO 14260, PI: D. Deming), which together cover a wavelength range of 0.8-1.7~$\mu$m. These observations revealed prominent H$_2$O absorption features. Combining the new HST/WFC3 observations with prior ones with VLT, HST/STIS and Spitzer, \citet{Wakeford2018} report an enriched atmospheric metallicity constraint of $151^{+48}_{-46} \times$~solar using a retrieval framework assuming chemical equilibrium. The median value corresponds to log-mixing ratios of $\sim-1.3$, $\sim-3.6$ and $\sim-4.8$ for H$_2$O, Na and K, respectively. By contrast, \citet{Tsiaras2018}, using an alternative reduction of the same HST/WFC3 G141 observations, find a significantly lower H$_2$O log-mixing ratio of $-5.94 \pm 0.61$. 

 This disparity in the inferred atmospheric composition of WASP-39~b persisted through subsequent analyses. Retrievals carried out by \citet{Pinhas2019} using the WFC3 G141 observations reduced by \citet{Tsiaras2018} along with HST/STIS and Spitzer data found an H$_2$O log-mixing ratio corresponding to a sub-solar metallicity, at $-4.07^{+0.72}_{-0.78}$, but Na and K log-mixing ratios corresponding to a super-solar metallicity, at $-3.86^{+1.31}_{-1.36}$ and $-4.22^{+1.25}_{-1.12}$. \citet{Kirk2019} presented a combined transmission spectrum consisting of the HST/WFC3 observations presented by \citet{Wakeford2018} and Spitzer photometry in the NIR, while combining HST/STIS and VLT observations with new observations using the William Herschel Telescope in the optical. Using a retrieval framework assuming equilibrium chemistry, they obtain an atmospheric metallicity constraint of $282^{+65}_{-58} \times$~solar, at its median corresponding to H$_2$O, Na and K log-mixing ratios of -0.9, -3.3 and -4.5, respectively. Using this same combined dataset, \citet{Welbanks2019b} find that composition constraints are dependent on the choice of prior, obtaining log-mixing ratio estimates of $-0.65^{+0.14}_{-1.83}$, $-3.62^{+1.14}_{-2.69}$ and $-5.62^{+2.30}_{-2.05}$ for H$_2$O, Na and K with their canonical retrieval, and $-2.43^{+0.27}_{-0.24}$, $-6.17^{+0.50}_{-0.51}$ and $-7.24^{+0.71}_{-1.06}$ for a more constrained prior. More recently, \citet{Kawashima2021} analysing the combined HST/STIS, WFC3 and Spitzer observations presented by \citet{Sing2016}, \citet{Fischer2016} and \citet{Wakeford2018}, constrain an atmospheric metallicity consistent with solar to within 1-$\sigma$ when considering disequilibrium chemistry. They also report a moderately super-solar metallicity constraint, corresponding to an H$_2$O log-mixing ratio of $\sim$-2.6.

 The JWST observations of WASP-39~b used in the present study have been obtained with the Near Infrared Spectrograph (NIRSpec) PRISM \citep{Ferruit2012, Birkmann2014} over a single transit in July 2022 as part of the JWST Early Release Science (ERS) \citep{ERS2022}. Spanning a subset of the full NIRSpec PRISM $\sim$0.6-5~$\mu$m range, the new data shows absorption peaks that are significantly larger in size than those observed previously in the HST/WFC3 bandpass, corresponding to 6-8 vs 2-3 atmospheric scale heights. Several reductions of the same observations were presented, which are reported to be largely comparable but with small deviations, especially in the 3-4~$\mu$m range. For the sake of robustness, we consider two reductions, based on their level of agreement over the Spitzer 3.6~$\mu$m bandpass, in order to assess the effect different reduction pipelines may have on the retrieved atmospheric properties. Specifically, we use the data obtained using the Eureka and Tiberius pipelines. Both the Eureka and Tiberius pipelines give rise to observations which, when binned to the Spitzer 3.6~$\mu$m bandpass, are at a higher transit depth than that observed by Spitzer itself. The Tiberius pipeline value is consistent with the Spitzer point to within 1-$\sigma$, while the Eureka value lies at $\sim$2-$\sigma$ of the Spitzer point and between those of the tshirt and FIREFLy pipelines. The 3-5~$\mu$m JWST data obtained with the Tiberius pipeline that are used in the present study are shown in figure \ref{fig:bestfit}, along with prior observations with HST/WFC3.

\section{Methods}\label{sec:methods}

We retrieve the atmospheric properties of WASP-39~b from the spectroscopic observations described in section \ref{sec:observations} using a variant of the AURA retrieval framework \citep{Pinhas2018}. The forward model computes radiative transfer in a plane-parallel atmosphere in transmission geometry. The model assumes hydrostatic equilibrium and local thermodynamic equilibrium in a H$_2$-rich atmosphere. The pressure-temperature (P-T) profile and uniformly-distributed volume mixing ratios of the chemical absorbers are free parameters in the model. In this work, we additionally retrieve the properties of Mie scattering aerosols, as discussed below. We also consider the conventional parametric cloud/haze prescription in AURA, for reference, as well as the effect of stellar heterogeneities. The parametric atmospheric model is coupled to a Bayesian inference and parametric estimation routine based on the Nested Sampling algorithm, implemented via the \textsc{PyMultiNest} package \citep{Feroz2009, Buchner2014, Feroz2019}.

In order to consider the spectral contributions of aerosols the model includes extinction from Mie scattering particles in the planetary atmosphere. Using the approach in \citet{Pinhas2017} we explore a range of possible condensate species that can be prevalent in irradiated giant exoplanets, e.g., MgSiO$_3$, Na$_2$S, MnS, ZnS, SiO$_2$, Al$_2$O$_3$, FeO, Fe$_2$O$_3$, TiO$_2$, NaCl and Mg$_2$SiO$_4$, based on data from \citet{Wakeford2015, Pinhas2017}. The extinction cross sections are computed following Mie theory \citep{Bohren1983}. We assume a modified gamma distribution for the aerosol particle sizes \citet{Deirmendjian1969}, with the modal particle size, $r_\mathrm{c}$, of the distribution, being a free parameter in the model. We additionally consider the vertical extent of the aerosol layer, described by the relative scale height of the aerosols, $h_\mathrm{c} = \frac{H_\mathrm{c}}{H}$, where $H_\mathrm{c}$ is the aerosol scale height and $H$ is the atmospheric scale height. $h_\mathrm{c}$ is another free parameter in the model with values ranging from 0 to 1. An $h_\mathrm{c}$ value of 1 implies that aerosols have a constant mixing ratio with altitude, while a value of 0 corresponds to no aerosols being present in the observable atmosphere. We incorporate this in our model as an exponential decrease in the aerosol mixing ratio with altitude:

\begin{equation}
X_{i}(z) = X_{i,0}~\mathrm{exp}\left[- \frac{(n-1)z }{H(z)} \right],
\label{eqn:Hc}
\end{equation}

where $X_i$ denotes the mixing ratio of the $i^\mathrm{th}$ aerosol species, $H(z)$ is the local atmospheric scale height $\frac{k_\mathrm{B} T}{\mu g}$ at an altitude $z$, and $n = \frac{1}{h_\mathrm{c}}$. Our model also accounts for an inhomogeneous coverage of the terminator atmosphere by aerosols, whose coverage fraction, $f_\mathrm{c}$, is a third free parameter.

Our aerosol model can include an arbitrary number of aerosol species. The mixing ratio of each of the aerosol species is a separate free parameter. For the retrievals we consider in this work, the modal particle size, vertical extent and fractional coverage parameters are universal, applying to all aerosol species in the model.

In light of the JWST observations probing a novel part of the spectrum and the high precision of observations, we carry out a staged retrieval approach. We begin by considering a maximal set of gaseous and Mie scattering aerosol species. This maximal model considers opacity contributions from a large number of gaseous chemical species. It also includes Mie scattering arising from inhomogeneous coverage of the terminator atmosphere by aerosols of MgSiO$_3$, Na$_2$S, MnS, ZnS, SiO$_2$, Al$_2$O$_3$, FeO, Fe$_2$O$_3$, TiO$_2$, NaCl and Mg$_2$SiO$_4$ \citep{Wakeford2015, Pinhas2017}.

We then consider a reduced canonical set of parameters, based on initial indications by our maximal retrieval and chemical expectations, which we use for all retrieval cases presented in this work. The final set of gaseous chemical species included in the present canonical model comprises of H$_2$O, CO, CO$_2$, H$_2$S, SO$_2$, CH$_4$, NH$_3$, HCN, C$_2$H$_2$. We additionally include opacity contributions arising from H$_2$-H$_2$ and H$_2$-He collision-induced absorption \citep{Richard2012}, as well as ZnS  \citep{Querry1987} and MgSiO$_3$ \citep{Dorschner1995} aerosols. Our choice for these two aerosol species is driven by both thermochemical expectations for the condensates based on the terminator temperature \citep{Morley2013} and indicative constraints obtained with our maximal model retrievals. The absorption cross-sections for the gaseous species are derived following \citet{Gandhi2017}, using line lists of H$_2$O, CO and CO$_2$ from \citet{Rothman2010} and \citet{Li2015}, CH$_4$ from \citet{Yurchenko2014}, NH$_3$ from \citet{Yurchenko2011}, HCN from \citet{Harris2006} and \citet{Barber2014}, C$_2$H$_2$ from \citet{Chubb2020} SO$_2$ from \citet{Underwood2016} and H$_2$S from \citet{Azzam2016} and \citet{Chubb2018}.

Our canonical atmospheric model has a total of 21 free parameters. The first 9 correspond to the individual log-mixing ratios of the gaseous chemical species listed above. Another two free parameters describe the log-mixing ratios of ZnS and MgSiO$_3$ aerosols and another 3 describe their fractional coverage, modal particle size and vertical extent, as described above. The terminator temperature profile is modelled by 6 parameters using the parametrisation of \citet{Madhusudhan2009}. The last free parameter for our canonical model is the planet radius, $R_\mathrm{P}$, defined at a nominal reference pressure of 0.1~bar. For retrievals on combined JWST and HST/WFC3 observations, we additionally retrieved for a linear offset between the two datasets.

We use log-uniform priors between $10^{-12}$-$10^{-0.3}$ for the mixing ratios of gaseous species, and between $10^{-30}$-$10^{-6}$ for the mixing ratios of MgSiO$_3$ and ZnS aerosols. We set the prior for the modal particle size, $r_\mathrm{c}$ to a log-uniform distribution ranging between 1~nm and 1$\mu$m and both the $f_\mathrm{c}$ and $h_\mathrm{c}$ priors are uniform between 0-1. The prior for the temperature at the top of the atmosphere, T$_\mathrm{0}$, is also uniformly distributed between 300-1600~K.

For completeness, we also explore retrievals including the effects of stellar heterogeneities as well as a more traditional parametric approach to model clouds/hazes, instead of Mie scattering by aerosols, as pursued by default in AURA. For the parametric clouds/hazes, we use a four-parameter combination of inhomogeneous grey opacity clouds and modified Rayleigh-like hazes \citep{Macdonald2017,Pinhas2018}. We incorporate stellar heterogeneities in the model following \citep{Rackham2017} as described in \citet{Pinhas2018}. This model involves three free parameters, describing the fractional surface coverage of heterogeneities, their overall effective temperature, and the temperature of the pristine photosphere.

 \begin{figure*}
    \centering

    \includegraphics[width=\textwidth]{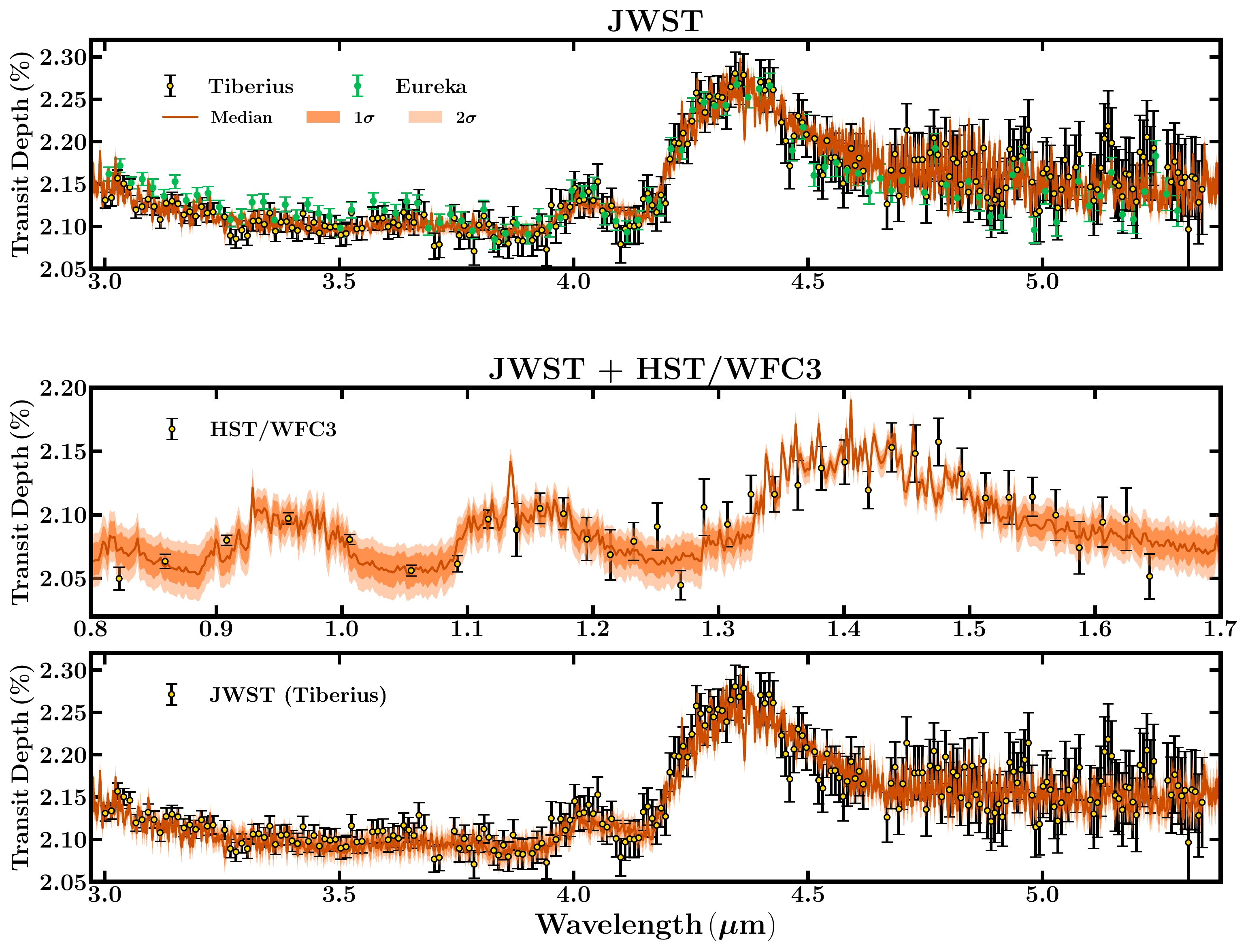}
    \caption{ Retrieved spectral fits obtained for two of the retrievals considered in this work, using our canonical model described in section \ref{sec:methods}. The top panel shows the retrieved spectral fit for JWST NIRSpec PRISM 3-5 $\mu$m observations reduced with the Tiberius pipeline (data shown in black errorbars with yellow circles). Also shown are the same observations reduced with the Eureka pipeline (in green). The lower two panels show different wavelength regions of the retrieved spectral fit to the combination of HST/WFC3 observations (0.8-1.7 $\mu$m) and JWST/NIRSpec PRISM observations reduced with Tiberius; see section~\ref{sec:observations}. In all three panels, the darkest orange line denotes the median retrieved spectrum while the two lighter orange regions denote the corresponding 1- and 2-$\sigma$ contours.}
    \label{fig:spectra}
\end{figure*}

%\vspace{0.5cm}
\section{Results}\label{sec:wasp-39b}

We now proceed to investigate the performance of atmospheric retrievals on a JWST spectrum of WASP-39~b. We first consider the JWST/NIRSpec PRISM observations on their own, examining the constraints such observations can lead to. We also assess how the retrieved atmospheric constraints vary due to minor differences between reduction pipelines, by considering JWST observations reduced by both Tiberius and Eureka. We then present our findings from joint retrievals on both the JWST observations over the 3-5~$\mu$m range and prior HST/WFC3 data. In doing so we establish the complementarity between JWST and HST/WFC3. We once again carry this out for data obtained with both the Tiberius and Eureka reduction pipelines, examining the differences between the resulting atmospheric constraints and how they vary with differences in our retrieved atmospheric model. Our retrieved constraints are summarised in table \ref{tab:constraints}.

\begin{deluxetable*}{l|cccccc}
\tablecaption{Retrieved atmospheric parameters for WASP-39~b.  \label{tab:constraints}}
\tablecolumns{8}

\tablehead{
\colhead{Case} & \colhead{\hspace{0.23cm} $\mathrm{log}(X_\mathrm{H_2O})$}\hspace{0.23cm} &  \colhead{\hspace{0.23cm} $\mathrm{log}(X_\mathrm{CO_2})$}\hspace{0.23cm}  & \colhead{\hspace{0.23cm} $\mathrm{log}(X_\mathrm{SO_2})$}\hspace{0.23cm} & \colhead{\hspace{0.23cm} $\mathrm{log}(X_\mathrm{CO})$}\hspace{0.23cm} & \colhead{\hspace{0.23cm} $\mathrm{log}(X_\mathrm{H_2S})$}\hspace{0.23cm} & \colhead{\hspace{0.23cm} $T_0 / \mathrm{K}$}\hspace{0.23cm}  }
\startdata
\hline
\multicolumn{7}{c}{Canonical Retrieval Model}\\
\hline
JWST Tiberius & $-4.85^{+0.38}_{-0.35}$  & $-6.28^{+0.38}_{-0.31}$ &  $-7.01^{+0.23}_{-0.20}$  & $-4.25^{+0.39}_{-0.35}$ & $-5.32^{+0.36}_{-0.42}$ & $862^{+64}_{-63}$ \\
JWST Eureka  & $-3.29^{+0.59}_{-0.56}$ & $-5.11^{+0.63}_{-0.54}$ & $-6.40^{+0.39}_{-0.35}$ & $-4.17^{+0.61}_{-0.61}$ & $-4.11^{+0.49}_{-0.46}$ & $738^{+54}_{-55}$ \\
Tiberius + HST/WFC3  & $-3.27^{+0.26}_{-0.24}$ & $-4.52^{+0.36}_{-0.30}$ & $-5.94^{+0.22}_{-0.19}$ & $-2.58^{+0.51}_{-0.50}$ & $-4.01^{+0.27}_{-0.24}$ & $757^{+40}_{-43}$  \\
Eureka + HST/WFC3 & $-3.28^{+0.33}_{-0.27}$ & $-4.57^{+0.51}_{-0.38}$ & $-6.31^{+0.25}_{-0.24}$ & $-3.61^{+0.37}_{-0.40}$ & $-4.17^{+0.29}_{-0.26}$ & $666^{+53}_{-72}$ \\
\hline
\hline
\multicolumn{7}{c}{Other Retrieval Models}\\
\hline
Tiberius + HST/WFC3, Parametric Cl./Hz. & $-3.69^{+0.31}_{-0.25}$ & $-4.75^{+0.41}_{-0.39}$  & $-6.21^{+0.24}_{-0.23}$ & $-2.40^{+0.47}_{-0.45}$ & $-4.49^{+0.31}_{-0.25}$ & $758^{+63}_{-61}$ \\
Eureka + HST/WFC3, Parametric Cl./Hz. & $-3.14^{+0.34}_{-0.31}$ & $-4.49^{+0.41}_{-0.41}$ & $-6.32^{+0.30}_{-0.29}$ & $-3.62^{+0.48}_{-0.62}$ & $-4.49^{+0.41}_{-0.41}$ & $683^{+52}_{-44}$ \\
Tiberius + HST/WFC3, Stellar Het. & $-3.44^{+0.27}_{-0.26}$ & $-4.65^{+0.34}_{-0.33}$ & $-6.05^{+0.22}_{-0.21}$ & $-2.85^{+0.42}_{-0.44}$ & $-4.19^{+0.26}_{-0.25}$ & $763^{+49}_{-51}$ \\
Eureka + HST/WFC3, Stellar Het. & $-3.34^{+0.31}_{-0.27}$ & $-4.58^{+0.41}_{-0.34}$ & $-6.29^{+0.24}_{-0.24}$ & $-3.58^{+0.35}_{-0.36}$ & $-4.09^{+0.24}_{-0.26}$ & $656^{+53}_{-48}$ \\
\enddata 
\tablecomments{The table shows the retrieved log-mixing ratios of molecules with notable detection significances along with the temperature at the top of the model atmosphere. The top four rows show the retrievals using our canonical model, with the top two obtained with JWST NIRSpec 3-5 $\mu$m data alone and the remaining two with a combination of JWST and HST data. We consider JWST data reported using two pipelines, Tiberius and Eureka, as discussed in section \ref{sec:wasp-39b}. The bottom four rows show constraints on the JWST+HST data obtained with two other retrieval considerations: (a) replacing the Mie scattering aerosols with a conventional parametric cloud/haze prescription, and (b) including stellar heterogeneities, as described in section \ref{sec:methods}.}

\end{deluxetable*}

\vspace{-0.8cm}
\subsection{Retrievals with JWST Data}\label{subsec:jwst_only}

We first focus on the recently-released JWST observations of WASP-39~b. As discussed in section \ref{sec:methods}, we analyse the observations with a staged retrieval approach. We begin by considering a maximal set of chemical species, including gases and  Mie scattering aerosols. We then consider a reduced set of chemical absorbers, based on physical plausibility and their initial constraints obtained by our maximal retrieval. Our reduced set of chemical species comprising our canonical model consists of H$_2$O, CO, CH$_4$, HCN, H$_2$S and SO$_2$, as well as ZnS and MgSiO$_3$ aerosols. \footnote{During the preparation of this work, we learned about  the independent inference of SO$_2$ using the same data \citep{Rustamkulov2022}. } 

We consider data obtained with the Tiberius and Eureka pipelines presented by \citet{ERS2022}, which produce somewhat different transit depths when binned over the Spitzer 3.6~$\mu$m photometric band. Specifically, the Tiberius pipeline produces data that are the closest to the Spitzer 3.6~$\mu$m transit depth measurement, with all other pipelines including Eureka which yields transit depths that are more than 1-$\sigma$ higher than the Spitzer value.

The retrieved spectral fit to the JWST observations with our canonical model is shown in figure \ref{fig:spectra}. The observations display a highly prominent absorption peak at 4.3~$\mu$m that has been previously attributed to CO$_2$ \citep{ERS2022}, as well as a smaller absorption feature at 4.0~$\mu$m. Moreover, the spectrum trends upwards at shorter wavelengths. For both reduction pipelines, our retrievals produce good fits to the data.

 \begin{figure*}
    
    \includegraphics[width=\textwidth]{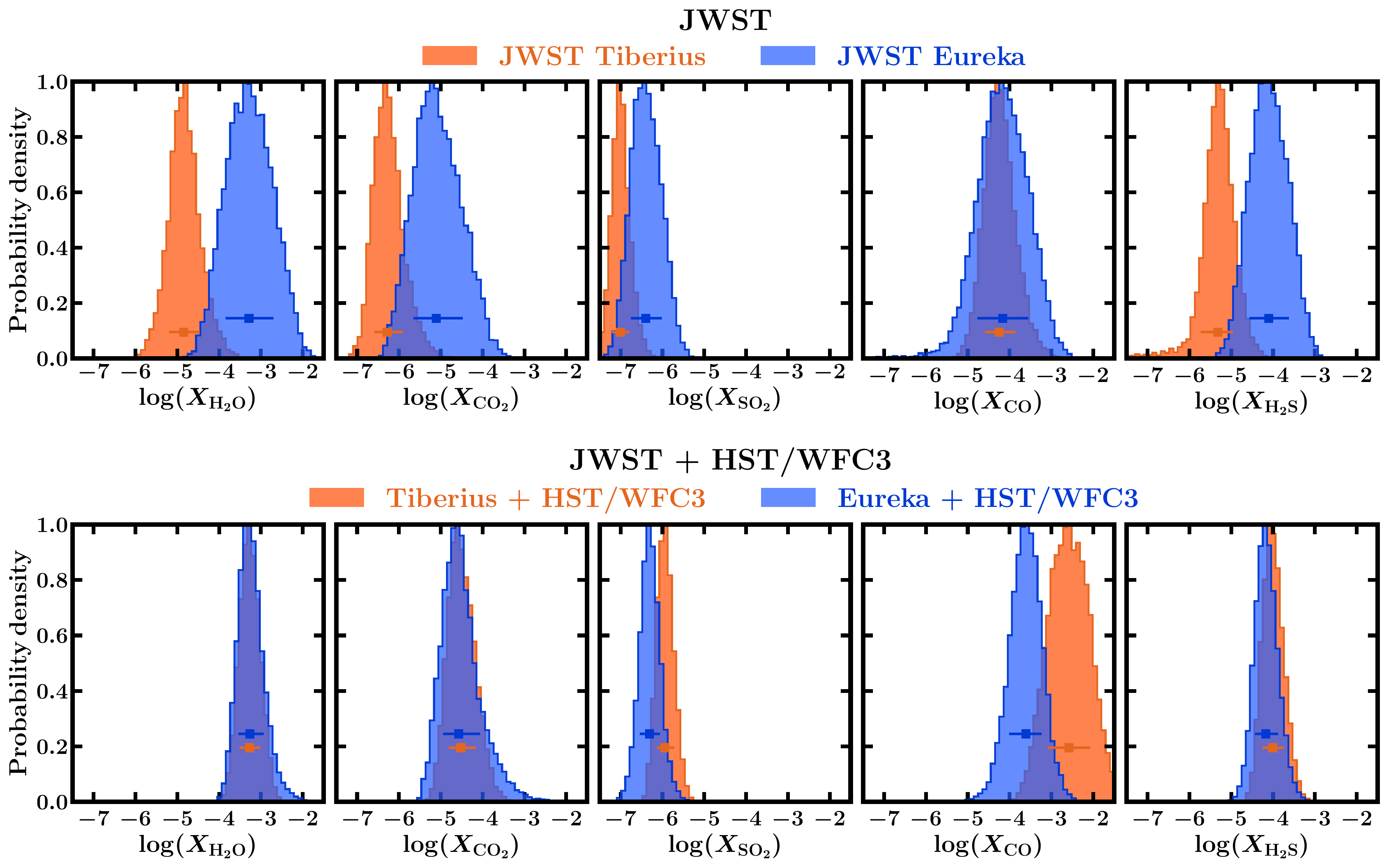}
    \centering
    \caption{Posterior distributions of retrieved molecular abundances. Top: Posteriors obtained with JWST NIRSpec PRISM 3-5~$\mu$m data reduced by the Tiberius and Eureka pipelines \citep{ERS2022}. Bottom: Posteriors with the same two JWST spectra combined with HST/WFC3 data \citep{Wakeford2018}. From left to right, the panels show the posteriors for log-mixing ratios of H$_2$O, CO$_2$, SO$_2$, CO and H$_2$S. Horizontal errorbars denote the retrieved median and 1-$\sigma$ interval for the posterior of corresponding colour.}
    \label{fig:posteriors}
\end{figure*}

\subsubsection{Tiberius Reduction}
\label{subsubsec:tiberius_only}

We begin by considering the data obtained via the Tiberius reduction pipeline. As shown in figure \ref{fig:spectra}, our retrieval obtains a good fit to the two significant absorption features at 4.0 and 4.3~$\mu$m, as well as the trend of increasing transit depth towards lower wavelengths. Moreover, the retrieval also fits smaller features within that trend. We confirm that the larger peak at 4.3~$\mu$m is due to CO$_2$ and the smaller peak at 4.0~$\mu$m  due to SO$_2$, as reported previously \citep{ERS2022, Rustamkulov2022}.

Our retrievals obtain constraints for the log-mixing ratios of CO$_2$, SO$_2$, H$_2$O, H$_2$S and CO. Driven by the very prominent CO$_2$ absorption peak at 4.3~$\mu$m, we constrain the log-mixing ratio of CO$_2$ to $-6.28^{+0.38}_{-0.31}$. Additionally, our retrieval attributes the smaller absorption peak seen at 4.0~$\mu$m to SO$_2$, constraining its log-mixing ratio to $-7.01^{+0.23}_{-0.20}$. CO is invoked to explain the data redward of the CO$_2$ feature, with a log-mixing ratio of $-4.25^{+0.39}_{-0.35}$. Additionally, H$_2$O and H$_2$S are constrained to log-mixing ratios of $-4.85^{+0.38}_{-0.35}$ and $-5.32^{+0.36}_{-0.42}$, respectively, and are used to fit the spectrum below 4.0~$\mu$m. The posterior distributions retrieved for each of these molecules are shown in figure \ref{fig:posteriors}.

In addition to the above mixing ratio constraints, our retrieval obtains a P-T profile that is consistent with an isotherm to within 1-$\sigma$, constraining T$_0$, the temperature at the top of the model atmosphere to $862^{+64}_{-63}$~K. Additionally, this retrieval does not obtain any constraints for the properties of Mie scattering aerosols. The posterior distributions for the log-mixing ratios of MgSiO$_3$ and ZnS which are largely unconstrained, with that of MgSiO$_3$ displaying a somewhat prominent peak at $\sim$-10. The posterior distributions for the modal particle size, fractional terminator coverage and vertical extent are also unconstrained.

We carry out additional retrievals to assess the detection significance for each of the constrained molecules presented above. We do this by performing a Bayesian model comparison between our canonical retrieval model and one without the molecule in question \citep{Pinhas2018}. We find that both SO$_2$ and, particularly, CO$_2$ are confidently detected. In the case of CO$_2$, the model including it is preferred at a $\sim$16-$\sigma$ level, while the inclusion of SO$_2$ in the model is favoured at a $\sim$4-$\sigma$ level. This is consistent with the fact that both molecules present significant absorption features in the observed wavelength range, and their exclusion therefore significantly deteriorates the achievable fit. All other molecules are retrieved with a lower detection significance. We find that H$_2$O, which was previously detected with HST/WFC3 observations, and CO, which was undetected before the advent of JWST, are both preferred at a $\sim$3-$\sigma$ level. Additionally, H$_2$S is marginally preferred at a $\sim$2-$\sigma$ level.

 We therefore find that while the detection significance of each molecule varies significantly, they are all retrieved with roughly similar precision, e.g. both CO$_2$ and H$_2$S are constrained with a precision of 0.4~dex, despite CO$_2$ having an extremely high detection significance while the H$_2$S is only marginally preferred. As such, the precision with which the abundance of a chemical species is estimated is not necessarily indicative of how robustly it is detected.

\subsubsection{Eureka Reduction}

We now consider retrievals carried out on the 3-5~$\mu$m JWST data obtained with the Eureka reduction pipeline. This dataset is more deviant from the Spitzer 3.6~$\mu$m channel datapoint than that from the Tiberius pipeline, with the resulting averaged transit depth lying $\sim$2~$\sigma$ higher than the Spitzer point. It is representative of multiple data reductions presented by \citet{ERS2022}.

With this dataset, our retrievals once again provide abundance constraints for several molecules. These include CO$_2$, at a log-mixing ratio of $-5.11^{+0.63}_{-0.54}$, as well as SO$_2$ at $-6.40^{+0.39}_{-0.35}$, which the retrieval invokes to explain the feature at 4.0~$\mu$m, similarly to our findings with the Tiberius reduction data. The retrieval also constrains the mixing ratios of H$_2$O, CO and H$_2$S to $-3.29^{+0.59}_{-0.56}$, $-4.17^{+0.61}_{-0.61}$ and $-4.11^{+0.49}_{-0.46}$, respectively. The retrieval additionally obtains posterior distributions for CH$_4$ and HCN which are notably peaked at log-mixing ratio values of $\sim$-7 and $\sim$-6, respectively, but have significant probability density extending to the lower end of the prior range. As such neither constitutes a precise or robust constraint.

The retrieval does not obtain any constraints for the properties of our included ZnS and MgSiO$_3$ aerosols. The mixing ratios for both aerosol species remain unconstrained, as were the posteriors for the fractional cloud coverage, particle size and vertical extent. Higher aerosol mixing ratios coincide with smaller particle sizes, which together result in negligible spectral contributions. We also once again retrieve a P-T profile that is consistent with an isotherm to within 1-$\sigma$. Our retrieval constrains T$_0$, the temperature at the top of the atmosphere, to $738^{+54}_{-55}$~K.

As before, we perform Bayesian model comparisons to assess the degree of model preference for including each molecule with abundance constraints. We find a very significant model preference towards the inclusion of CO$_2$ at $\gtrsim$20-$\sigma$, while SO$_2$ and H$_2$O are preferred at a $\sim$4-$\sigma$ level. The detection significances for CO and H$_2$S are once again not as high as those obtained for CO$_2$ and SO$_2$, both of which are preferred $\sim$3-$\sigma$ level.

The inclusion of CO in the model is favoured at a $\sim$3-$\sigma$ level. Lastly, the inclusion of H$_2$S is preferred at a $\sim$4-$\sigma$ level. As with our prior retrieval on data from the Tiberius pipeline, we find that the precision with which the mixing ratio of each species is constrained is not indicative of how confidently it is detected.

\subsubsection{Comparison of Retrieved Constraints}

Both reduction pipelines lead to mixing ratio constraints with precisions below one dex. Additionally, both lead to extremely confident detection significances for CO$_2$ as well as a slightly less confident but still robust detection of SO$_2$. They also both result in moderate model preferences in favour of H$_2$O and CO. 

Despite the above, we find significant differences in atmospheric parameters retrieved with the two datasets. Most notably, despite their precision, the retrieved abundance constraints for CO$_2$, SO$_2$, H$_2$O and CO from the two datasets are not consistent to within 1-$\sigma$, in some cases differing by 1~dex or more. This indicates that each dataset leads retrievals to a different spectral baseline. As a result, retrievals then invoke different amplitudes of spectral features in order to explain the data. Another significant difference is in the detection significance of H$_2$S, with Eureka leading to a relatively robust detection of 4-$\sigma$ while Tiberius leads to only a tentative indication of its presence, with a 2-$\sigma$ detection significance. We additionally find other pipeline-specific features, in the form of peaked, but largely unconstrained posterior distributions for HCN and CH$_4$ obtained with the Eureka pipeline data but not with the Tiberius data.

We also find differences between the retrieved temperature profiles. While both retrievals obtain P-T profiles that are consistent with an isotherm, they lie more than 2-$\sigma$ away from each other. This is likely another consequence of the two datasets leading retrievals to different spectral baselines. 

Thanks to the extreme precision of the JWST observations, we therefore find that pipeline-specific features can lead to significant differences in retrieved quantities. As such, we find that while both retrievals lead to molecular detections of key species, they can lead to significantly different mixing ratio estimates. As such, the significance with which a chemical species is detected may not always indicate an accurate abundance estimate, when considering spectra in the $\sim$3-5~$\mu$m range alone.

%\vspace{-0.2cm}
\subsection{Retrievals on Combined JWST and HST Observations}\label{subsec:combined_obs}

 \begin{figure*}
    \includegraphics[width=\textwidth]{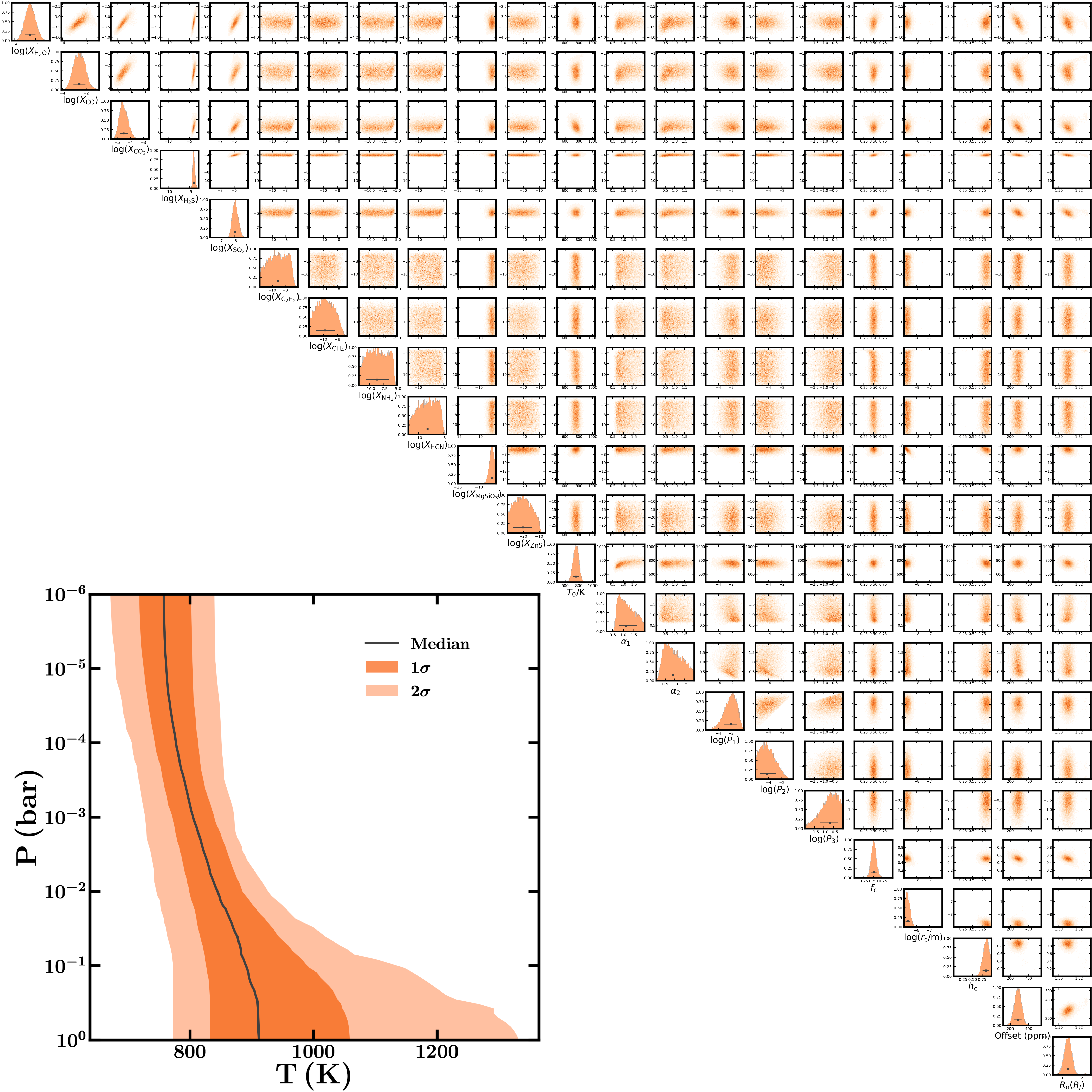}
    \centering
    \caption{Full posterior distribution from the retrieval using the JWST/NIRSpec PRISM 3-5 $\mu$m spectrum, reduced with the Tiberius pipeline, combined with HST/WFC3 data (0.8-1.7 $\mu$m). The model parameters correspond to the canonical atmospheric model described in section \ref{sec:methods}. Horizontal errorbars denote the median and 1-$\sigma$ interval for each retrieved parameter. Also shown is the retrieved P-T profile, with the black line denoting the median retrieved profile, while darker and lighter orange contours indicate the 1- and 2-$\sigma$ intervals.}
    \label{fig:JWST_cornerplot}
\end{figure*}

 We now examine how atmospheric constraints retrieved from JWST observations  in the $\sim$3-5~$\mu$m are affected when we additionally include observations at shorter wavelengths obtained with HST/WFC3. This allows us to assess the complementarity of JWST observations over the 3-5~$\mu$m range, which form a substantial part of JWST Cycle 1 programs, with HST spectra at shorter  wavelengths (0.8-1.7 $\mu$m). As before, we consider JWST spectra obtained with the Tiberius and Eureka pipelines and in both cases combine them with HST/WFC3 G102 and G141 observations in the $\sim$0.8-1.7$\mu$m range presented by \citet{Wakeford2018}. We retrieve with the same canonical atmospheric model as in prior sections, which is described in detail in section \ref{sec:methods}. As noted there, we additionally retrieve for a vertical linear offset between the JWST and HST/WFC3 datasets.

\vspace{+0.5cm}
\subsubsection{Tiberius Reduction and HST/WFC3}
\label{subsubsubsec:tiberius+wfc3}

We first consider adding HST/WFC3 observations to $\sim$3-5~$\mu$m JWST/NIRSpec PRISM data reduced with the Tiberius pipeline. Our retrieval once again achieves a good fit to the JWST/NIRSpec observations, while also finding a good fit to the HST/WFC3 data. The best-fit spectrum, along with the corresponding scale heights of different features is shown in figure \ref{fig:bestfit}. The individual molecular opacity contributions to the best-fit spectrum are shown in figure \ref{fig:contributions} in the Appendix. Additionally, the retrieved median spectral fit and corresponding 1- and 2-$\sigma$ contours are shown in figure \ref{fig:spectra}. The retrieval invokes H$_2$O to explain the HST/WFC3 data as expected, while the JWST observations are explained with CO$_2$, SO$_2$ H$_2$S and CO. Notably, the size of the CO$_2$ feature is significantly larger ($\sim$8 scale heights) than that of H$_2$O in the HST/WFC3 band ($\sim$2-3 scale heights).

The retrieved atmospheric constraints from this combined dataset retrieval are notably different to those obtained from the JWST data alone, with the increased spectral coverage at shorter wavelengths leading the retrieval to better constrain the spectral baseline. Specifically, the retrieved log-mixing ratios for CO$_2$ and SO$_2$ are now $-4.52^{+0.36}_{-0.30}$ and $-5.94^{+0.22}_{-0.19}$, while that of H$_2$O is constrained to $-3.27^{+0.26}_{-0.24}$. Additionally, CO and H$_2$S are constrained to log-mixing ratios of $-2.58^{+0.51}_{-0.50}$ and $-4.01^{+0.27}_{-0.24}$. Compared to the estimates obtained from our retrieval on Tiberius-derived data alone, the present constraints are all higher by $\sim$1~dex or more. The complete posterior distribution is shown in figure \ref{fig:JWST_cornerplot}.

The increase in wavelength coverage also allows our retrieval to constrain the broad wavelength contributions from Mie scattering MgSiO$_3$ aerosols. Specifically, we obtain a log-mixing ratio constraint of $-6.99^{+0.55}_{-0.65}$ for MgSiO$_3$ particles, with a modal particle size of $\mathrm{log}(r_\mathrm{c} / \mu\mathrm{m}) = -2.71^{+0.20}_{-0.17}$. These particles are found to be extended up to high altitudes, with a relative scale height, $h_\mathrm{c}$ of $0.85^{+0.08}_{-0.9}$ and occupying roughly half of the terminator atmosphere, with a coverage fraction constrained to $51^{+7}_{-7} \%$. Meanwhile, we also find an upper limit for the mixing ratio of ZnS particles of -9.81 at 99\% confidence. This indicates that the data are best fit by spectral features specific to MgSiO$_3$.

The retrieved spectral contributions from the MgSiO$_3$ aerosol particles are such that there is significant opacity contributions over the HST/WFC3 wavelength range, while the 3-5~$\mu$m range covered by the NIRSpec PRISM observations lie mostly within an opacity window. As such, it is likely that retrievals on these JWST observations alone are unable to distinguish between a cloud-free case and one with most of the observations lying within an opacity window of a partially cloudy atmosphere.

Our retrieval also finds a P-T profile that is not consistent with an isotherm to within 1-$\sigma$, as shown in figure \ref{fig:JWST_cornerplot}. Specifically, our retrieval constrains T$_0$, the temperature at the top of the atmosphere to $757^{+40}_{-43}$~K. The constrained P-T profile then increases in temperature at higher pressures, with the median profile reaching a temperature of $\sim$900~K at a pressure of 1~bar. 

We once again carry out a Bayesian model comparison to assess the detection significance for each molecule. We find that the inclusion of CO$_2$ is again very strongly preferred, at a $\sim$20~$\sigma$ level. H$_2$O is now also strongly preferred, at a $\sim$~13~$\sigma$ level, thanks to the addition of HST/WFC3 observations which encompass strong H$_2$O absorption features. SO$_2$ meanwhile is preferred at a $\sim$4~$\sigma$ level while now the inclusion of H$_2$S is also preferred at a $\sim$4~$\sigma$ level. Lastly, the inclusion of CO is preferred at $\sim$3~$\sigma$ level. H$_2$S in particular is now more strongly preferred than when retrieving on JWST data alone in section \ref{subsubsec:tiberius_only}.

We therefore find that the inclusion of HST/WFC3 observations are highly informative to retrievals on JWST observations in the $\sim$3-5~$\mu$m range. This is evident in the significantly higher retrieved abundance constraints relative to those obtained with the same JWST data alone in section \ref{subsubsec:tiberius_only}, as well as their increased precision and detection significances. Moreover, we find that combined HST/WFC3 and JWST observations over the 3-5~$\mu$m range can, in principle, lead to constraints for the physical properties of atmospheric aerosols, as well as the terminator's temperature structure.

\vspace{-0.15cm}
\subsubsection{Eureka Reduction and HST/WFC3}

We now consider pairing the JWST/NIRSpec 3-5~$\mu$m data obtained with the Eureka pipeline with HST/WFC3 G141 and G102 observations. Unlike the retrievals on data from the Tiberius pipeline described above, we find that combining HST/WFC3 data with observations reduced with Eureka does not lead to significant changes in the retrieved atmospheric properties. Specifically, our retrieval constrains the log-mixing ratio of CO$_2$ to $-4.57^{+0.51}_{-0.38}$, SO$_2$ to $-6.31^{+0.25}_{-0.24}$, H$_2$O to $-3.28^{+0.33}_{-0.27}$, CO to $-3.61^{+0.37}_{-0.40}$ and H$_2$S to $-4.17^{+0.29}_{-0.26}$. These constraints are generally more precise than those obtained with the corresponding JWST/NIRSpec data alone. 

We find tentative indications of spectral contributions from Mie-scattering aerosols using this dataset. Specifically, we find an upper limit for the mixing ratio of ZnS of -6.65 at 99\% confidence, which corresponds to significant spectral contributions. The same is true for the constraints obtained for $\mathrm{log}(r_\mathrm{c} / \mu\mathrm{m})$, $f_\mathrm{c}$ and $H_\mathrm{c}$, which have 99\% confidence upper limits of 0.96, -6.15 and 0.97. This indicates that the data do not preclude significant spectral contributions from aerosols. Meanwhile, the posterior for the mixing ratio of MgSiO$_3$ aerosols is unconstrained.

The retrieval finds an atmospheric P-T profile that is consistent with an isotherm to within 1-$\sigma$. Specifically, it constrains the T$_0$, the temperature at the top of the atmosphere to $666^{+53}_{-72}$~K. Notably, this temperature is more than 1-$\sigma$ away from that obtained with only JWST/NIRSpec data.

As with other retrievals, we find that the inclusion of CO$_2$ in our retrieved atmospheric model is very strongly preferred, with a detection significance of $\sim$25~$\sigma$. H$_2$O is also preferred at a lower but still highly confident $\sim$12~$\sigma$ level. Meanwhile, SO$_2$, H$_2$S and CO are all preferred at a $\sim$3~$\sigma$ level.

We therefore find that the addition of HST/WFC3 observations to the Eureka pipeline data once again affects the retrieved atmospheric properties. In addition to improving the precision of all abundance constraints, including those with no features in the HST/WFC3 band, it also leads to different results for aerosol parameters.

\vspace{-0.05cm}
\subsubsection{Comparison of Retrieved Constraints}

We find that the retrieved mixing ratio values for H$_2$O, CO$_2$, SO$_2$ and H$_2$S all now agree to within 1-$\sigma$ between the two JWST pipelines. This is a notable difference with our prior retrievals on JWST data alone, which differed by 1~dex or more. On combining HST/WFC3 data with JWST data from the Tiberius pipeline, we find that there is a significant change in the retrieved abundance constraints, in some cases increasing by more than 1~dex relative to those obtained from JWST observations alone. Meanwhile, adding HST/WFC3 observations to data from the Eureka pipeline does not result in as significant a shift in retrieved mixing ratios.

Despite the general agreement between the retrieved mixing ratio values, differences still persist between the two JWST datasets. Most notably, the retrieved mixing ratios of CO differ by more than 1-$\sigma$. Such pipeline-specific constraints persist when we consider an expanded retrieval model, such as one including KOH and NaOH, which may be present based on prior Na and K constraints. In this case, we find that the HST/WFC3 + Eureka dataset leads to a preference for KOH rather then H$_2$S, while retrieving on HST/WFC3 + Tiberius still leads to a preference for H$_2$S over KOH.

Secondly, we find different reduction pipelines also lead to different constraints for aerosols present in the atmosphere. In the case of our retrievals on the HST/WFC3 + Tiberius dataset, we obtain precise constraints on sub-micron MgSiO$_3$ aerosols covering roughly half the terminator atmosphere with an effectively full vertical atmospheric extent. The retrieval also specifically invoked MgSiO$_3$ as opposed to ZnS, which is also a part of our canonical model. Meanwhile, the retrieval on HST/WFC3 + Eureka data instead leads to only tentative constraints for ZnS aerosols and none for MgSiO$_3$.

Lastly, we also find different retrieved temperature profiles. Retrieving on HST/WFC3 + Tiberius data, we find a non-isothermal P-T profile, which may also be consistent with that obtained with Tiberius pipeline data alone, depending on the specific altitude probed by the retrieval. Meanwhile, the HST/WFC3 + Eureka dataset leads to a P-T profile that is consistent with an isotherm. It is also inconsistent with the temperature obtained with Eureka pipeline data alone, as well as the temperature obtained with HST/WFC3 + Tiberius. The temperature constraint is relevant for interpreting the retrieved atmospheric composition and understanding the physical and chemical processes giving rise to them. We also conclude that while precise transmission spectra over a wide wavelength range can in principle lead to constraints for the terminator's temperature structure, such constraints are sensitive to minor variations in the data, such as those introduced by differences in reduction pipelines.

We also consider how our findings are affected by changes to our model, described in section \ref{sec:methods}, the results of which are summarised in table \ref{tab:constraints}. On including the effects of stellar heterogeneities in our model, we find that our retrieved abundance constraints are consistent with those obtained with our canonical model. We also consider the impact of using a parametric cloud/haze prescription rather than physically-motivated Mie-scattering aerosols. In this case, we find that our retrieved abundances are different by up to $\sim$0.5~dex, underscoring the need for a physically motivated aerosol model in order to obtain accurate abundance constraints in the JWST era.

\section{Summary and Discussion}\label{sec:discussion}

We use the first JWST observations of an exoplanet transmission spectrum in the 3-5 $\mu$m range to obtain early insights into atmospheric retrievals that are possible in the JWST era. This spectral range is particularly important to investigate, considering the large allocation of JWST time in Cycle 1 for exoplanet spectroscopy using NIRSpec observations in the $\sim$3-5$\mu$m range. We focus on the JWST NIRSpec PRISM observations of the hot Saturn WASP-39b, observed as part of the JWST ERS program \citep{ERS2022}, arguably one of the most promising exoplanets for transmission spectroscopy. Our goal is to investigate (a) the nature of atmospheric constraints that are possible with high-precision JWST data in the $\sim$3-5~$\mu$m range, (b) modeling requirements for atmospheric retrievals with such data, (c) how sensitive the retrieved atmospheric properties are to differences in spectra from different reduction pipelines, and (d) how such data may be complemented with observations at shorter wavelengths.

\subsection{Key Lessons}

We identify the following key lessons for atmospheric retrievals of transiting exoplanets using JWST transmission spectra, particularly in the 3-5 $\mu$m range.

\begin{itemize}

\item The $\sim$3-5~$\mu$m range provides an important avenue for molecular detections, as it encompasses windows in extinction cross section for several aerosols. Consequently, molecular opacity within this range can give rise to highly prominent spectral features. JWST observations over the $\sim$3-5~$\mu$m may therefore lead to confident molecular detections even for potentially cloudy atmospheres with low-amplitude ($\lesssim$2-3 scale height) spectral features in the HST/WFC3 hand ($\sim$0.8-1.7~$\mu$m) \citep[e.g.][]{Stevenson2016a, Fu2017, Crossfield2017}. Our results add to studies using simulated JWST observations \citep{Wakeford2015, Pinhas2017, Mai2019, Lacy2020a}.

\item In order for retrievals to obtain accurate abundance estimates with JWST observations in the $\sim$3-5~$\mu$m range, complementary observations may be needed at shorter wavelengths. This allows retrievals to constrain the spectral baseline needed for accurate and precise estimates of molecular abundances. Specifically, we find that complementary observations with HST/WFC3 can be highly effective for this goal, resulting in changes in the retrieved abundances by up to one dex in some cases, compared to retrievals using $\sim$3-5~$\mu$m spectra alone.

\item JWST observations  over a wide wavelength range can in principle constrain the presence of clouds/hazes, as well as their specific nature, e.g. particle composition and size. This has been previously considered with simulated JWST observations \citep[e.g.][]{Benneke2013, Wakeford2015, Pinhas2017, Mai2019, Lacy2020a} and with HST-era observations \citep{Benneke2019}. It is therefore important that atmospheric retrievals have the capability to properly treat the complex spectral signatures that aerosols can contribute to a planet's transmission spectrum.

\item As expected from prior theoretical works, JWST transmission spectra in the near-infrared can provide precise abundance constraints for several prominent molecules in exoplanetary atmospheres\citep[e.g.][]{Beichman2014, Greene2016, Howe2017, Kalirai2018, Bean2018}. For the present case, we retrieve  abundance constraints with a precision of $\sim$0.3-0.5~dex for prominent molecules, even with the relatively limited spectral range considered. Combining this data with other JWST observations can lead to even more precise abundance constraints.

\item Small differences in JWST spectra, such as those arising due to different reduction pipelines, can have a notable impact on the retrieved constraints and detections, particularly forless prominent species. This is thanks to the high precision that JWST observations can have, especially for giant exoplanets. For the case of the WASP-39~b observations we consider, which are for a single transit, we find that differences in spectra arising from differences between reduction pipelines can affect which species are detected at a 2-3$\sigma$ level. Therefore, robust data reduction pipelines are needed in order to converge on accurate chemical detections.

\item A high detection significance for a particular chemical species does not indicate that its abundance is constrained accurately. For instance, retrieving on JWST/NIRSpec PRISM observations obtained with the Tiberius pipeline leads to a $\sim$25$\sigma$ detection significance. However, on supplementing these observations with HST/WFC3 G141 and G102 data, the retrieved mixing ratio estimate is 1~dex higher, as the retrieval finds a different spectral baseline.

\item The precision with which the abundance of a chemical species is constrained is not indicative of how confidently it is detected. For instance, using JWST NIRSpec PRISM observations reduced with the Tiberius pipeline, we constrain the mixing ratios of CO$_2$ and H$_2$S with a precision of $\lesssim$0.4~dex. While CO$_2$ is detected with extremely high confidence, however, H$_2$S is not, with our retrievals finding only a tentative 2-$\sigma$ model preference for its inclusion. It is therefore essential that future studies assess the robustness of their findings by carrying out a full Bayesian model comparison for each chemical species that may be detected.

\item The spectral ranges of JWST instruments allow for the detection of chemical species that have not been hitherto considered in exoplanet retrievals, and potentially chemical processes that may have not been anticipated. In the present example of WASP-39~b, we independently confirm the presence of SO$_2$ reported by \citet{Rustamkulov2022} and \citet{Alderson2022}, while also finding tentative indications of H$_2$S. We therefore find that retrieval frameworks must be open to diverse chemistry if they are to be capable of fully taking advantage of JWST observations.

\end{itemize}

\subsection{ The Atmosphere of WASP-39~b}
\label{subsec:composition_inferences}

 \begin{figure}
    \includegraphics[width=0.99\columnwidth]{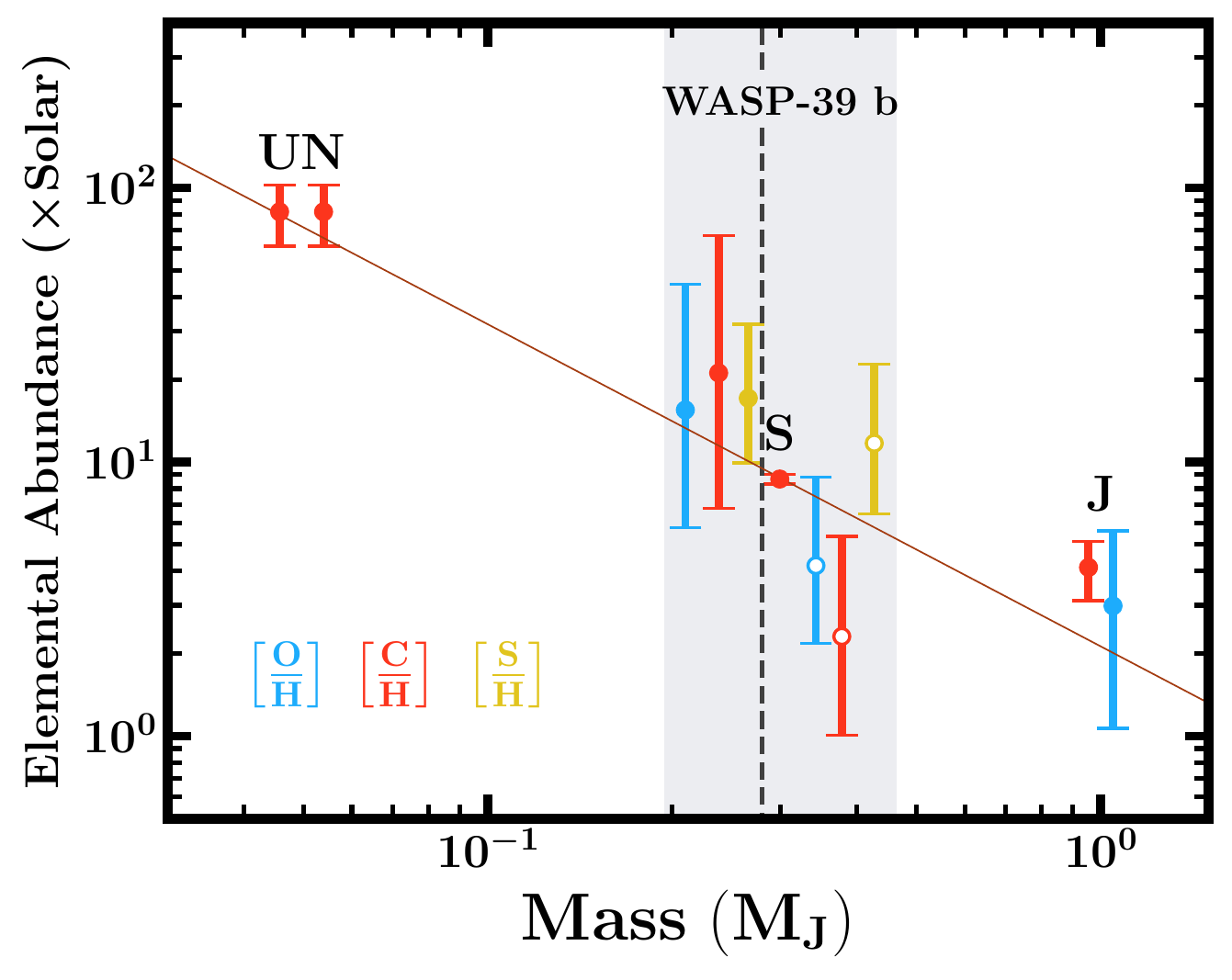}
    \centering
    \caption{Initial constraints on the elemental abundances in the atmosphere of WASP-39~b. The O/H (blue), C/H (red) and S/H (yellow) ratios are derived from the molecular abundances retrieved from JWST data in the 3-5 $\mu$m range combined with HST/WFC3 data (0.8-1.7 $\mu$m). The elemental abundances are shown relative to solar values \citep{Asplund2021}. The dashed vertical line denotes the planet mass (0.28~M$_\mathrm{J}$). Closed and open circles within the gray shaded region refer to retrievals using JWST data from the Tiberius and Eureka pipelines, respectively, offset horizontally for clarity. The C/H abundances for solar system giant planets \citep{Atreya2022} along with a linear fit (in brown) and the O/H in Jupiter from Juno \citep{Li2020} are shown for reference.}
\label{fig:mass_metallicity}
\end{figure}

In light of the above considerations, we present initial constraints on the atmospheric properties of WASP-39~b. We confirm the presence of CO$_2$, CO and SO$_2$ reported by \citet{ERS2022} and \citet{Rustamkulov2022, Alderson2022}. We additionally find tentative indications for the presence of H$_2$S, with detection sigificances between $\sim$2-4~$\sigma$ across our retrievals. Our abundance estimate of SO$_2$ supports prior findings of disequilibrium chemistry at work in the atmosphere of WASP-39~b \citep{Tsai2022}. Additionally, the presence of H$_2$S suggested by our retrievals further supports this, as H$_2$S is expected to be the primary sulfur reservoir under chemical equilibrium deeper in the atmosphere \citep{Zahnle2009, Wang2017, Hobbs2021, Polman2022}, which photochemically reacts with H$_2$O in the upper atmosphere to form SO$_2$.

As our retrievals obtain abundance constraints for a range of oxygen-, carbon- and sulfur-bearing species, we can begin to probe the relative enrichment of each element in the atmosphere of WASP-39~b, as shown in figure \ref{fig:mass_metallicity}. Considering the constraints obtained with JWST observations reduced by the Tiberius pipeline combined with HST/WFC3 data, we find that the inferred O/H, C/H and S/H values correspond to $15^{+30}_{-10}\times$, $21^{+46}_{-14}\times$ and $17^{+15}_{-7}\times$~solar enrichments, respectively \citep{Asplund2021}. Such enhancements are consistent with the atmospheric metallicity of Saturn based on CH$_4$ measurements, at $8.67 \pm 0.35 \times$~solar, and with recent suggestions for WASP-39~b\citep{Alderson2022,Rustamkulov2022,Feinstein2022}. Meanwhile, the equivalent O/H, C/H and S/H inferences using JWST observations reduced with the Eureka pipeline and HST/WFC3 data correspond to  $4^{+6}_{-2}\times$, $2^{+4}_{-1}\times$ and $11^{+11}_{-7}\times$~solar enrichments. In this case, the oxygen and sulfur enrichments are still consistent with the metallicity of Saturn, while carbon is somewhat less enriched. We note that the uncertainties in our O/H and C/H estimates are primarily driven by the CO mixing ratio constraints, which are retrieved with precisions of $\sim$0.4-0.5~dex. Retrievals combining the present JWST/NIRSpec PRISM data with recently-published NIRSpec G395H observations\citep{Alderson2022} that also encompass the $\sim$4.8~$\mu$m CO feature may further refine the present estimates.

Beyond constraints for gaseous species, we also obtain tentative indications of non-gray opacity contributions from Mie-scattering aerosols. Specifically, our retrieval on JWST data reduced with the Tiberius pipeline and HST/WFC3 observations infers the presence of MgSiO$_3$ aerosols. These aerosols are found to have a modal particle size of $\sim$2~nm and cover $\sim$50\% of the planet's terminator. Meanwhile, retrieving on JWST data from Eureka and HST/WFC3 observations, we find tentative indications of ZnS aerosols instead. These compositional constraints are consistent with thermochemical expectations for condensing species \citep{Morley2013}. Our findings are also in agreement with preferences for non-grey spectral contributions from aerosols obtained with a forward model analysis of NIRISS observations \citep{Feinstein2022}.

Our retrievals also lead to constraints for the planet's terminator temperature. Using the Tiberius reduction of the JWST data along with HST/WFC3 data, we find tentative indications of a non-isothermal temperature, with a temperature of $757^{+40}_{-43}$~K at the top of the atmosphere. Meanwhile, JWST data from Eureka and HST/WFC3 observations instead lead to a P-T profile that is consistent with an isotherm, with a temperature of $666^{+53}_{-72}$~K. 

WASP-39~b and its ERS observations are set to allow an in-depth comparison between a solar-system planet and an extrasolar planet. Our results contribute an early step towards a next-generation comparative study between Saturn and an exoplanet that is a near-Saturn analogue in both mass and metallicity.

The results presented above are only initial constraints for the atmospheric properties of WASP-39~b, obtained with a relatively limited spectral range of JWST observations ($\sim$3-5~$\mu$m). Future retrievals with all available data ($\sim$0.6-5~$\mu$m) can significantly improve upon the constraints presented in this work. However, our study highlights the need for rigorous data reduction and retrieval considerations as well as a robust exploration of the available parameter space. The retrievals in this work involved $\sim$$10^9$ model evaluations to achieve that goal.

In addition to physically motivated aerosol considerations, future retrievals may also consider the effect of inhomogeneous temperature \citep{Nixon2022} and chemical \citep{Rocchetto2016} structures, as well as the impact of stellar heterogeneity aided by optical data \citep{Rackham2017,Pinhas2018} and other molecular opacity sources. While such model sophistication may not be applicable in all cases, especially for moderately irradiated planets like WASP-39~b, a holistic approach is recommended for accurate retrievals of atmospheric properties in the JWST era.

Our results provide a glimpse into the richness of atmospheric science that JWST is set to enable, thanks to its unprecedented sensitivity and wide spectral coverage that includes hitherto unexplored wavelengths. Our results also highlight the sophistication demanded of atmospheric models and retrievals, as well as the robustness required of reduction pipelines, if we are to rise to the challenge and make full use of the discovery opportunities that JWST presents. 

%\begin{acknowledgments}
This work was performed using resources provided by the Cambridge Service for Data Driven Discovery (CSD3) operated by the University of Cambridge Research Computing Service (\url{www.csd3.cam.ac.uk}), provided by Dell EMC and Intel using Tier-2 funding from the Engineering and Physical Sciences Research Council (capital grant EP/P020259/1), and DiRAC funding from the Science and Technology Facilities Council (\url{www.dirac.ac.uk}). This work uses data reduced from observations made with the NASA/ESA/CSA JWST, as part of The Transiting Exoplanet Community Early Release Science (ERS) Program (PI: N. Batalha). We thank NASA, ESA, CSA, STScI, the ERS team, and everyone whose efforts have contributed to JWST and exoplanet science with JWST.

We thank the anonymous referee for their valuable comments. SC thanks Anjali Piette for helpful discussion on Mie scattering. NM acknowledges support from the MERAC Foundation, Switzerland, and the UK Science and Technology Facilities Council (STFC). SG is grateful to Leiden Observatory at Leiden University for the award of the Oort Fellowship. 
%\end{acknowledgments}

\appendix

\begin{figure*}[h]
    \centering
    \includegraphics[width=\textwidth]{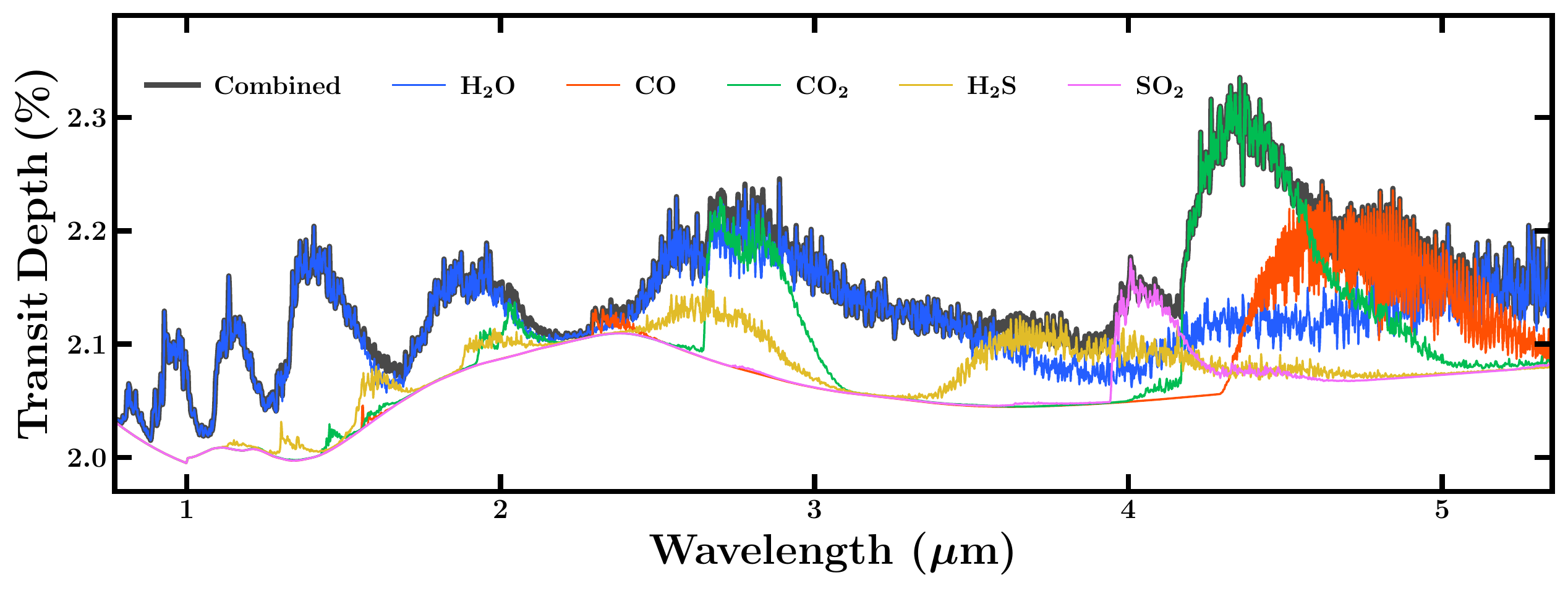}
    \caption{The spectral contributions from H$_2$O, CO, CO$_2$, SO$_2$ and H$_2$S to the best fit spectrum obtained from the retrieval on JWST observations reduced by the Tiberius pipeline combined with HST/WFC3 data. Also shown is the resulting spectrum from all molecular spectral contributions, corresponding to the aerosol-free model shown in figure \ref{fig:bestfit}. Spectral contributions from MgSiO$_3$ aerosols as shown in figure~\ref{fig:bestfit} are not included here, for visual clarity.}
    \label{fig:contributions}
\end{figure*}

Figure \ref{fig:contributions} shows the spectral contributions from gaseous species in our retrieved best fit spectrum, obtained for JWST/NIRSpec PRISM observations reduced with Tiberius and HST/WFC3 data. We note that this figure does not include spectral contributions from MgSiO$_3$ aerosols, for visual clarity. As such, the combined spectral contributions correspond to the blue curve in figure \ref{fig:bestfit}.

\bibliography{refs}{}
\bibliographystyle{aasjournal}

\end{document}